\documentclass[a4paper,fleqn,usenatbib,useAMS]{mnras}

\usepackage{graphicx,multicol} 
\usepackage{times}
\def\etal{\it et~al.}

\title[Drifting, Nulling and Mode changing in J1822$-$2256]{Subpulse Drifting, Nulling and Mode changing in PSR J1822$-$2256}
\author[Basu \& Mitra]{Rahul Basu$^{1}$, Dipanjan Mitra$^{2,3}$ \\
$^{1}$ Inter-University Centre for Astronomy and Astrophysics, Pune, 411007, India; rahulbasu.astro@gmail.com \\
$^{2}$ National Centre for Radio Astrophysics, Tata Institute of Fundamental Research, Pune 411007, India \\
$^{3}$ Janusz Gil Institute of Astronomy, University of Zielona G\'ora, ul. Szafrana 2, 65-516 Zielona G\'ora, Poland \\
}
\begin{document}



\maketitle

\label{firstpage}

\begin{abstract}
We report a detailed observational study of the single pulses from the pulsar 
J1822$-$2256. The pulsar shows the presence of subpulse drifting, nulling as 
well as multiple emission modes. During these observations the pulsar existed 
primarily in two modes; mode A with prominent drift bands and mode B which was 
more disorderly without any clear subpulse drifting. A third mode C was also 
seen for a short duration with a different drifting periodicity compared to 
mode A. The nulls were present throughout the observations but were more 
frequent during the disorderly B mode. The nulling also exhibited periodicity 
with a clear peak in the fluctuation spectra. Before the transition from mode A
to nulling the pulsar switched to a third drifting state with periodicity 
different from both mode A and C. The diversity seen in the single pulse 
behaviour of the pulsar J1822$-$2256 provides an unique window into the 
emission physics.
\end{abstract}


\section{Introduction}
\noindent
The single pulse sequences from pulsars, particularly the longer period 
pulsars, show a number of different phenomena like subpulse drifting, nulling 
and mode changing which are representative of the physical processes 
responsible for the radio emission. A general model has emerged which explains 
the generation of ultra-relativistic plasma in the pulsar magnetosphere leading
to coherent radio emission \citep{stu71,rud75,mit17b}. In this picture the 
plasma is believed to originate in the inner acceleration region (IAR) which is
characterised by large electric and non-dipolar magnetic fields. Sparking 
discharges take place due to pair production from $\gamma$-ray photons in the 
large magnetic fields of the IAR resulting in a non-stationary plasma flow. The
multi-component relativistic plasma clouds undergo nonlinear instabilities 
around heights of five hundred kilometers above the stellar surface to form 
charge separated solitons. The coherent radio emission arises as curvature 
radiation from the solitons moving in curved magnetic field lines \citep{ass98,
mel00,gil04,mit09}. The only source of periodicity in the single pulse dynamics
is due to the $\}bf{E}\times\bf{B}$ drift in the IAR which is believed to be 
responsible for subpulse drifting. The other periodic or quasi-periodic 
phenomena associated with pulsar emission, like nulling and mode changing, are 
difficult to explain using the above model. The periodic nulling is associated 
with empty sight line passes between the rotating subbeam system \citep{her07}.
However, in a recent work \citet{bas17} found that the empty sight line 
traverse was not adequate to explain the periodic nulling, particularly in 
pulsars with core components. It was reported by \citet{bas16,mit17} that the 
periodic amplitude modulation is very different from subpulse drifting. As a 
result \citet{bas17} suggested that the standard model of emission is no longer
adequate to explain the different phenomena seen in pulsars. An additional 
triggering mechanism from outside the IAR must periodically affect the plasma 
flow in the pulsar magnetosphere. In this context the coexistence of multiple 
drifting and non-drifting modes and nulling in the same pulsar is further 
indication of an external triggering mechanism. Therefore, characterising these
phenomena should serve as important inputs into understanding the origin of the
triggering mechanism.

There are only a handful of pulsars where multiple drifting states exist in 
addition to nulling and mode changing. The most well studied pulsar in this 
group is PSR B0031$-$07 where the drifting periodicity changes around the nulls
\citep{hug70,viv97,smi05,mcs17}. Similar coexistence of multiple subpulse 
drifting modes with nulling is seen in the pulsars B1918+19 \citep{han87,
ran13}, B1944+17 \citep{dei86,klo10}, B2319+60 \citep{wri81} and B2303+30 
\citep{red05}. In some of these pulsars the nulling is also periodic in nature
\citep{bas17}. Additionally, pulsars B1918+19 and B1944+17 have also been 
reported to have non-drifting modes. In the Meterwavelength Single-pulse 
Polarimetric Emission Survey \citep[][MSPES]{mit16,bas16,bas17} a possible 
existence of multiple drifting states in addition to nulling was reported for 
the pulsar J1822$-$2256. The subpulse drifting and nulling in this pulsar have 
also been reported by \citet{wel06,wel07,ser09,nai17}. However, the limited 
lengths of the observations as well as lower sensitivities of the single pulses
in these studies did not facilitate a proper characterisation of the emission 
properties. In this work we have carried out longer duration and more sensitive
observations to study the single pulse dynamics in the pulsar J1822$-$2256. We 
have characterised with greater accuracy the subpulse drifting, nulling and the 
mode changes seen in this pulsar. In addition we have also explored the 
physical processes in the pulsar magnetosphere that can give rise to these 
variations in the single pulse properties.

\section{Observations and Analysis}
\noindent
We have carried out extensive observations of the pulsar J1822$-$2256 using the
Giant Meterwave Radio Telescope (GMRT) located near Pune, India \citep{swa91}. 
The GMRT is an interferometric array consisting of thirty antennas each of 
forty five meter diameter, with fourteen antennas located within a central 
square kilometer area and the remaining sixteen antennas spread out along three 
arms in a Y-shaped array. We have used the Telescope in the phased array mode 
where the signals from different antennas were co-added. In order to reach 
sufficient sensitivity for single pulse studies we used approximately twenty 
antennas, including all the available central square antennas and the two 
nearest arm antennas. A phase calibrator was recorded at the start of the 
observation as well as every hour and appropriate ``phasing'' solutions were 
estimated to correct for temporal gain variations for each antenna. This 
resulted in phasing breaks between the recorded single pulse sequence. We 
observed the pulsar on 5$^{th}$ November 2015 for approximately three hours. 
The pulsar has a period of 1.87 seconds which ensured around 5700 single pulses
for these studies.

The observations recorded total intensity signals from the pulsar with the 
maximum frequency set at 339 MHz and spread over 33 MHz bandwidth. In contrast
the MSPES carried out full polarization studies with only 16 MHz bandwidth. The 
increased bandwidth resulted in higher sensitivity detections of the single 
pulses. However, the polarization information from the earlier observations was
also used to investigate the emission properties. The time resolution of the
observations was 491.52 microseconds. The known dispersion measure 
(121.20~pc~cm$^{-3}$) was used to correct for the temporal spread across the 
frequency band. Subsequently, the dispersion corrected signals were averaged 
across all frequencies to produce a series of total intensity measurements from
the pulsar for the entire observing duration. During the phasing breaks 
suitably weighted noise signals were inserted in between the pulsar signal to 
preserve continuity for fluctuation spectral studies. Finally, a two 
dimensional pulse stack with one axis along the pulse longitude, separated into
integral bins, and the other along the pulse number, was formed from the time 
series signals. Different analyses for identifying emission modes, measuring 
subpulse drifting features and nulling were carried out on the pulse stack.

As mentioned earlier due to the increased bandwidth the single pulses were more
prominent and we inspected them visually to identify the different emission 
modes. The subpulse drifting was characterised using the fluctuation spectral 
analysis \citep{bac73,bac75}. We used the Longitude Resolved Fluctuation 
Spectra (LRFS) where Fourier transforms across each longitude was carried out 
for a certain sequence of single pulses. The subpulse drifting periodicities 
were seen as frequency peaks in the fluctuation spectra. In addition to the 
the average LRFS studies we have also determined shorter duration LRFS 
corresponding to different emission modes. In the MSPES studies proper nulling 
analysis could not be carried out due to weaker sensitivities of the single 
pulses. In this work we have utilized the improved sensitivities to carry out a
detailed analysis of the null and burst pulses. The techniques used to 
characterize the nulling behaviour are detailed in \citet{bas17}. We 
established energy histograms for the on-pulse and off-pulse regions of the 
pulsar profile to identify the nulling fractions. Initially the null and burst 
pulses were identified using statistical boundaries. Subsequently, we visually 
inspected every null to eliminate any erroneous identification. The null and 
burst sequences were characterized by the respective null length and burst 
length histograms. Finally, a sequence of `0' and '1' was setup identifying the
null and burst pulses respectively. An FFT was carried out on this sequence to 
identify any periodicity associated with nulling.

\section{The Emission Modes and Subpulse Drifting}
\begin{figure*}
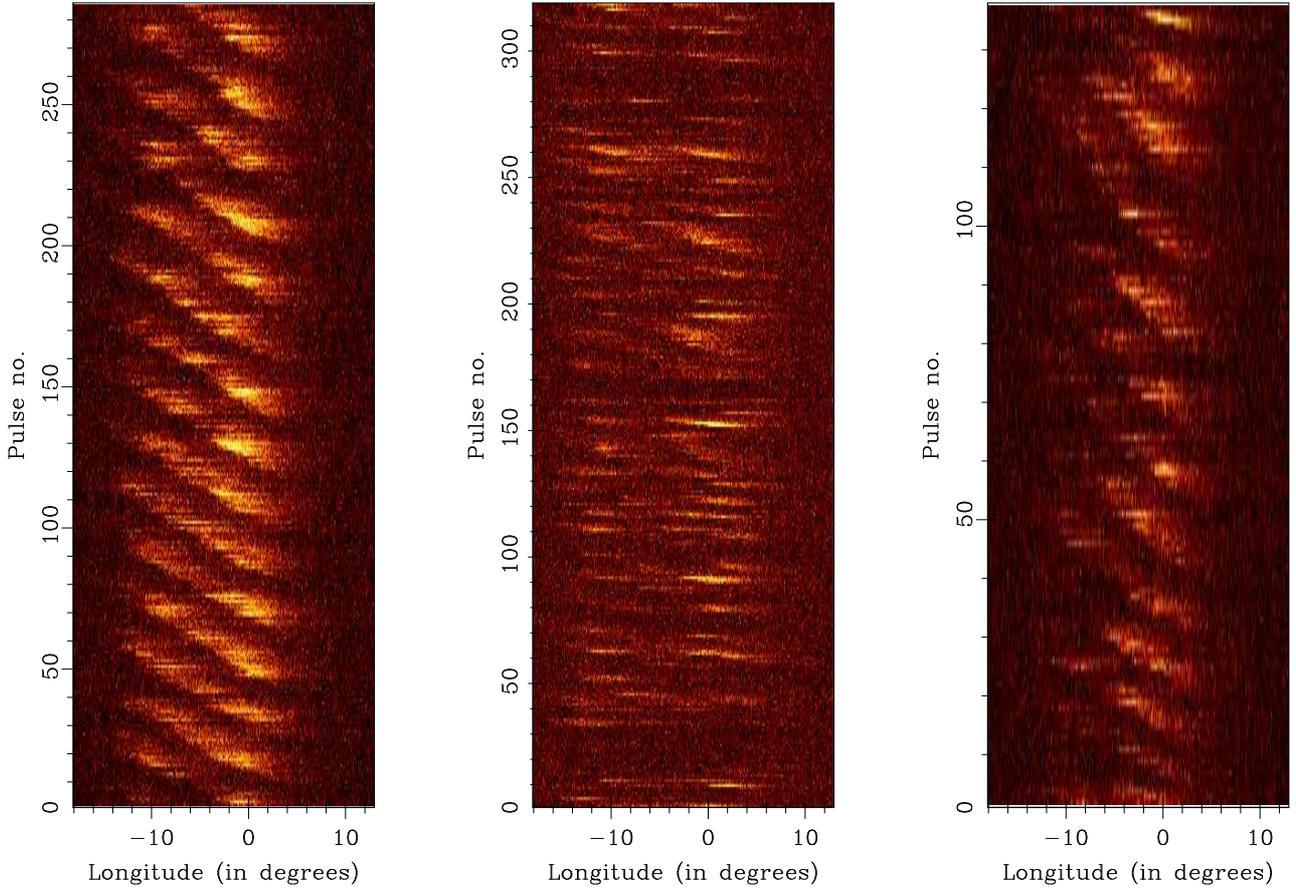

\begin{tabular}{@{}cr@{}cr@{}}
{\mbox{\includegraphics[scale=0.5,angle=0.]{modeAsngl.ps}}} &
\hspace{20px}
{\mbox{\includegraphics[scale=0.5,angle=0.]{modeBsngl.ps}}} &
\hspace{30px}
{\mbox{\includegraphics[scale=0.5,angle=0.]{modeCsngl.ps}}} \\
\end{tabular}
\caption{The figure shows the three different emission modes seen in the single
pulse sequence of PSR J1822$-$2256. The left panel shows a sequence of mode A
between pulse number 2679 and 2963 with regular drift bands. The central panel 
shows the single pulses from pulse number 3595 to 3912 and corresponds to mode
B. This mode does not exhibit any clear subpulse drifting and has more frequent 
nulls compared to mode A. The right panel shows single pulses between 3018 and 
3154 and is identified as mode C. The mode also shows distinct drift bands like
mode A but unlike mode A is more prominent towards the trailing edge of the 
pulse window. The drifting periodicity in mode C is also different from mode A.}
\label{fig_modesngl}
\end{figure*}

\begin{figure}
\includegraphics[scale=0.63,angle=0.]{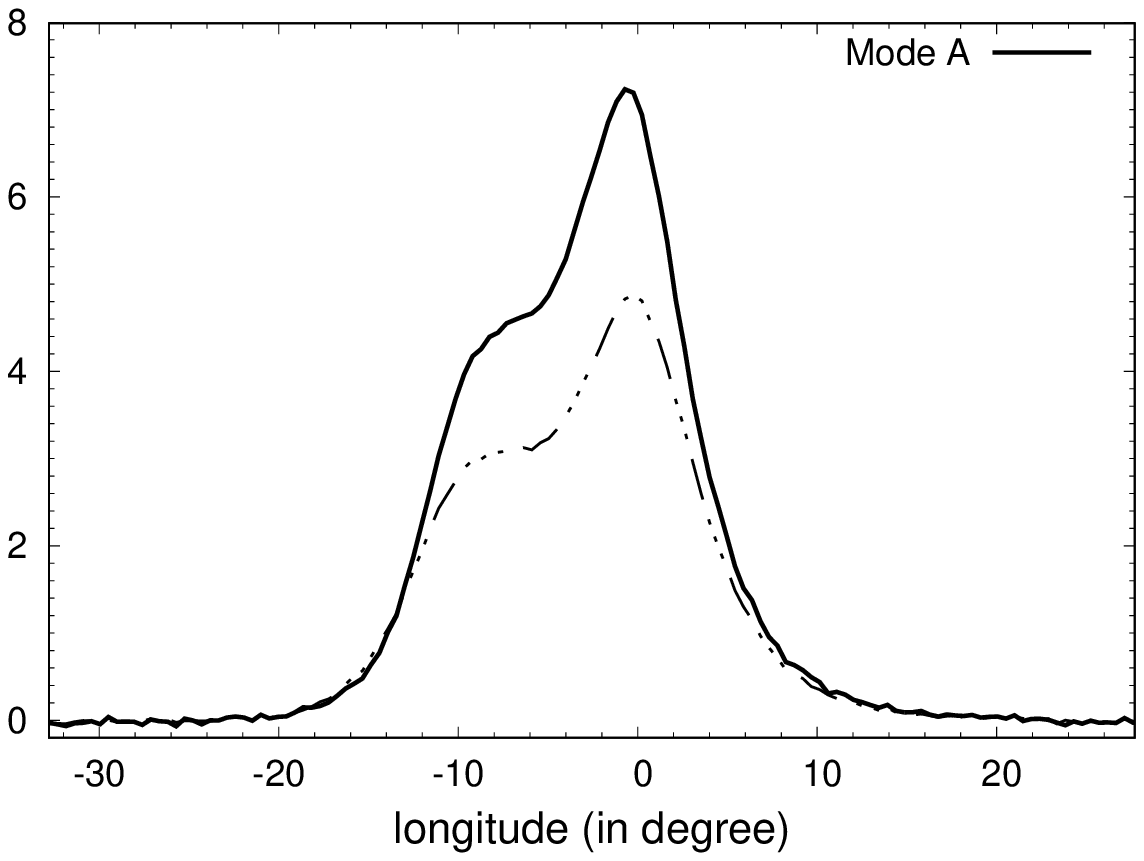} \\
\includegraphics[scale=0.63,angle=0.]{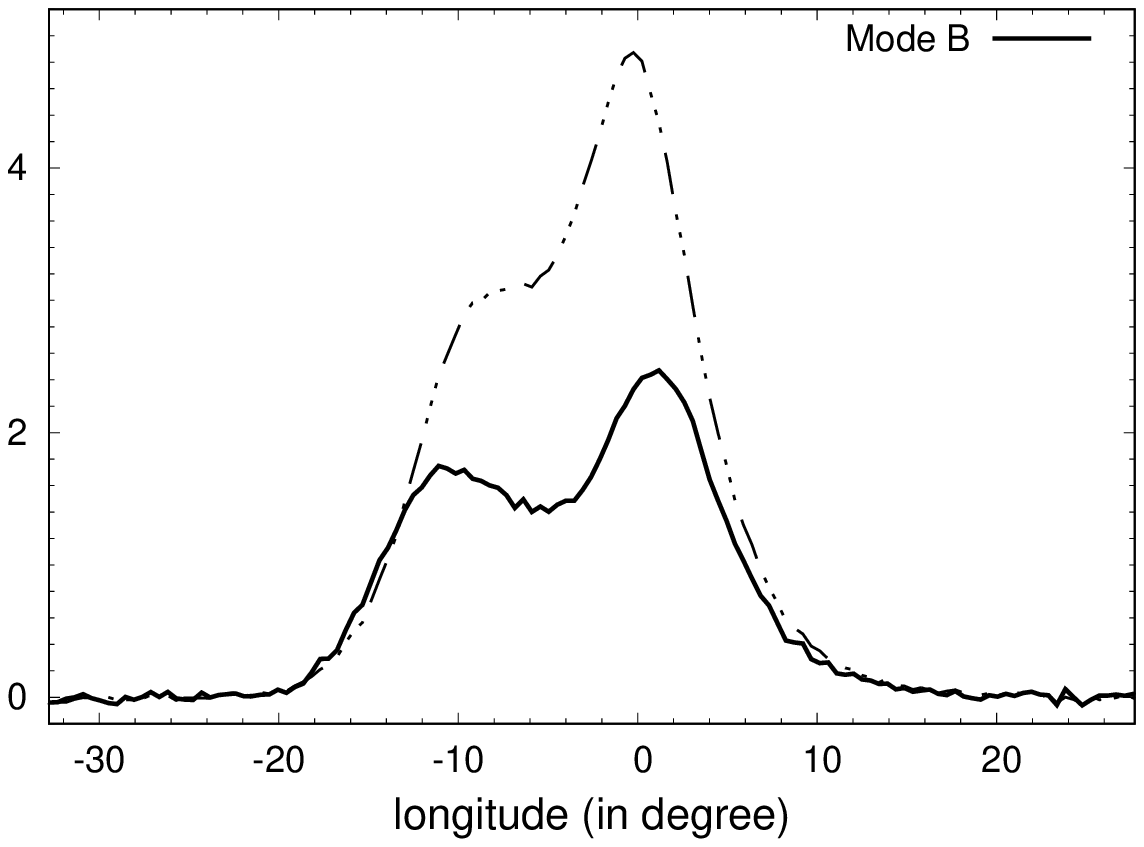} \\
\includegraphics[scale=0.63,angle=0.]{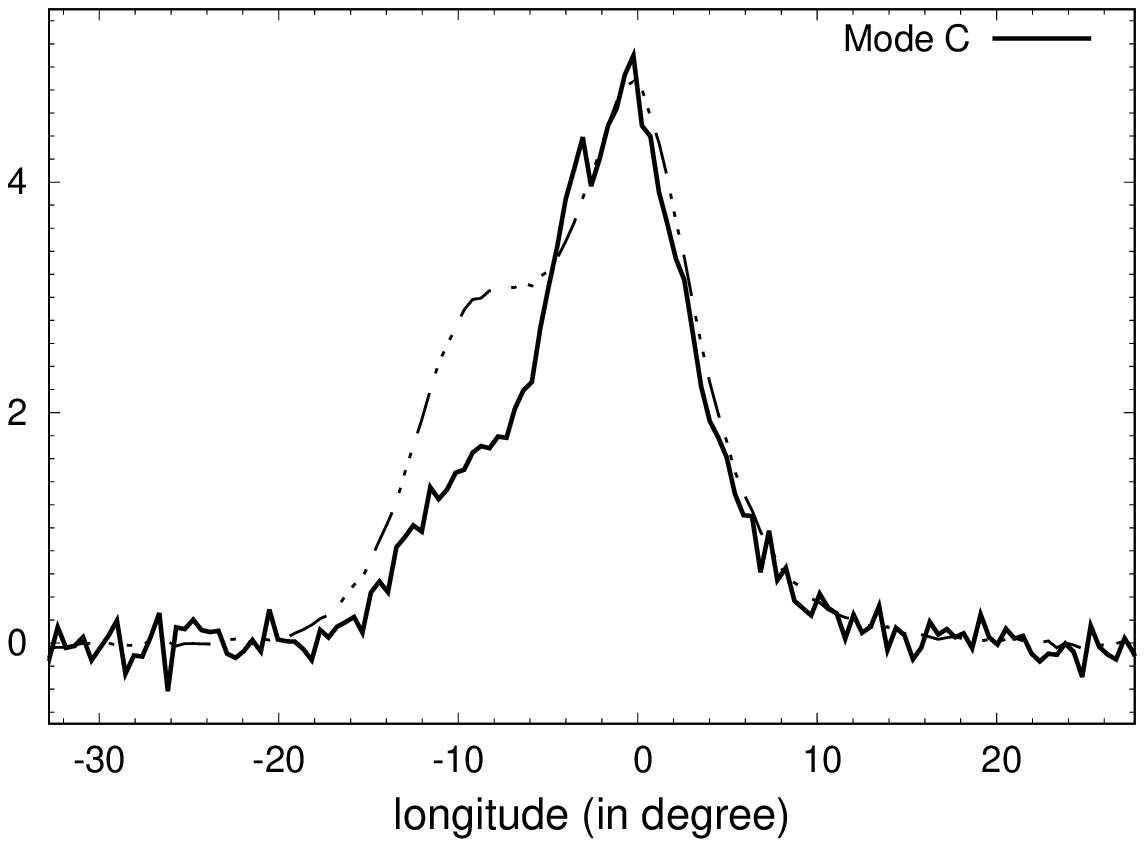} \\
\caption{The figure shows the variations in the profile shapes in the three 
distinct emission modes of the pulsar J1822$-$2256 superposed on the average 
profile (dot dashed line). The top panel corresponds to mode A with prominent 
drift bands across the pulse window and is brighter near the trailing side. The
middle panel plots the profile of the second mode B which is less bright and do
not show any clear drifting pattern. The bottom panel corresponds to mode C 
which also shows subpulse drifting with different characteristics compared to 
mode A. The profile indicates that the trailing edge is brighter in this mode.}
\label{fig_modeprof}
\end{figure}

\begin{figure}
\begin{center}
\includegraphics[scale=0.5,angle=0.]{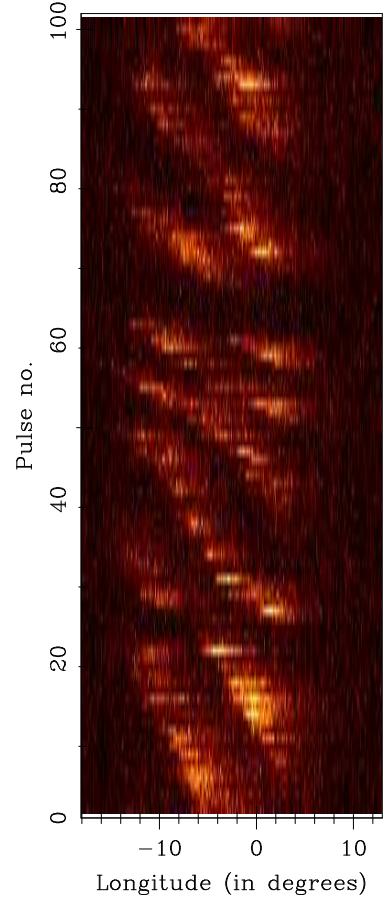}
\end{center}
\caption{The figure shows the distinct behaviour seen sometimes in mode A 
before the transition to the null state. The drifting nature changes before the 
onset of the nulls with the periodicity of subpulse drifting becoming much 
faster. The pulsar recovers to the original drifting state just after nulling.}
\label{fig_nulltransingl}
\end{figure}

We have identified three distinct emission modes in the pulsar J1822$-$2256 
based on its single pulse properties as shown in figure \ref{fig_modesngl}. The
primary distinguishing feature amongst the different modes was the nature of 
subpulse drifting. In the majority of the single pulse sequence we were able to
identify the modes by visual inspection. However, in some instances it proved 
difficult either due to mixing of the modes or reduction in intensity due to 
scintillation. The most prevalent emission mode was classified as mode A and 
showed prominent drift bands from the leading to the trailing edge of the pulse
window. The corresponding profile in figure \ref{fig_modeprof} (top panel) 
shows the trailing part to be brighter than the leading part in this mode. Mode
A was present for roughly 45 percent of the time during the observing duration. 
The average length of the mode was 82 periods with the minimum duration being 
28 periods and maximum duration being 288 periods. The pulsar transitioned 
frequently to the second mode B which was more disorderly and did not show any 
clear drift bands. The profile in figure \ref{fig_modeprof} (middle panel) 
resembles a double peaked structure much weaker than mode A. Mode B was 
somewhat less frequent than mode A and was seen in around 38 percent of the 
single pulses during these observations. The average modal length was 68 
periods, with minimum duration of 18 periods and maximum duration of 319 single
pulses. Finally, the least frequent but quite distinct mode C also showed 
prominent drift bands with drifting periodicity different from mode A (see 
table \ref{tabdrift}). In this mode the pulsar was brighter towards the 
trailing part of the profile window as seen in the profile shape (figure 
\ref{fig_modeprof}, bottom panel). The mode was the least frequent and only 
seen once for around 200 periods which correspond to 4 percent of the observing
duration. In addition to the the three distinct modes of emission the pulsar 
also showed nulling which was spread out throughout the observations but seen 
more frequently in the weaker B mode. Another different emission feature arose
sometimes within mode A before the transition from the drifting state to nulls.
The drifting property changed, as shown in figure \ref{fig_nulltransingl}, with
the periodicity of subpulse drifting becoming much shorter. However, this 
lasted for only a short duration each time and was not identified as a separate
mode. There was no clear ordering seen in the mode changing between different 
states. In some instances the pulsar changed from mode A to transition state 
and then to nulls followed by mode B. At other occasions the transition to mode 
B did not happen after the nulls and the pulsar reverted back to mode A. 

\subsection{Estimating Drifting properties}
\begin{table}
\resizebox{\hsize}{!}{
\begin{minipage}{80mm}
\caption{The Subpulse Drifting properties during the different emission modes 
of PSR J1822$-$2256.}
\centering
\begin{tabular}{cccccc}
\hline
 Mode & $f_p$ & FWHM & $S$ & $P_3$ & $P_2$ \\
      & (cy/$P$) & (cy/$P$) &   &  ($P$)  & (\degr) \\
\hline
 Mean & 0.058$\pm$0.010 & 0.023 & 17.2 & 17.1$\pm$2.9 & --- \\
      &    &    &    &    &  \\
   A  & 0.051$\pm$0.004 & 0.010 & 61.8 & 19.6$\pm$1.6 & 8.3$\pm$0.1 \\
      &    &    &    &    &  \\
Trans. (A) & 0.070$\pm$0.009 & 0.020 & 27.4 & 14.3$\pm$1.8 & 9.1$\pm$0.1 \\
      &    &    &    &    &  \\
   C  & 0.093$\pm$0.010 & 0.023 & 16.5 & 10.7$\pm$1.1 & 7.6$\pm$0.1 \\
      &    &    &    &    &  \\
\hline
\end{tabular}
\label{tabdrift}
\end{minipage}
}
\end{table}

\begin{figure}
\begin{center}
\includegraphics[scale=0.42,angle=0.]{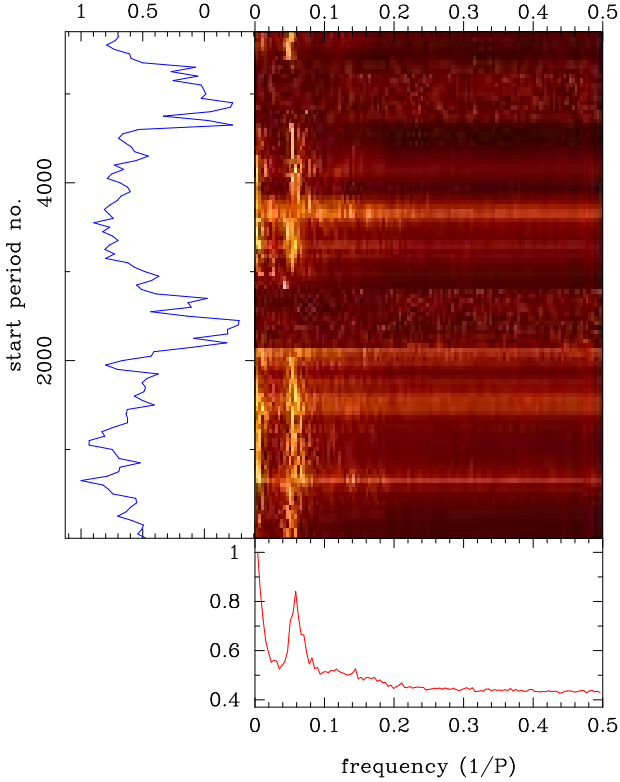} 
\end{center}
\caption{The figure shows the variation of the average Longitude Resolved 
Fluctuation spectra (LRFS) as a function of the start period. The LRFS is 
determined for 256 consecutive single pulses at a time. The starting point is 
shifted by 50 periods and the process is continued till the end of the 
observing duration. The average LRFS is calculated for each realization and is
represented in the colour plot as a function of the starting period. The 
average across the x-axis is shown on the left window while the average across 
the y-axis corresponding to the time average LRFS is shown on the bottom 
window. The time average LRFS exhibits two clear peak frequencies. The first 
peak at zero frequency can be attributed to nulling while the second peak 
corresponds to the subpulse drifting in the dominant mode A. The figure also 
shows the two intervals during phasing of the antennas where the corresponding 
LRFS has been replaced with noise.}
\label{fig_lrfsfull}
\end{figure}

\begin{figure}
\begin{center}
\includegraphics[scale=0.41,angle=0.]{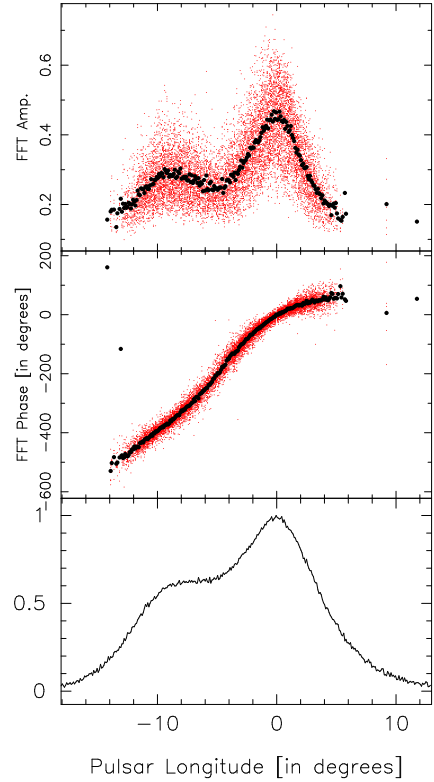}\\
\end{center}
\caption{The figure shows the variations of the peak amplitude (top window) 
and the phase (middle window) of the Longitude Resolved Fluctuation spectra 
(LRFS) across the pulse profile. The average profile shape is shown in the 
bottom window. The LRFS is determined for 256 consecutive single pulses at a 
time. The starting point is shifted by 50 periods and the process is continued 
till the end of the observing duration. In each case the peak amplitude 
represents the subpulse drifting in mode A. The spread of the peak amplitude 
and phase, represented by red points, corresponds to their variations with time 
while the black dots are the average values at each longitude. The amplitude 
variations resemble the profile shape. The phase variations are non-linear with
the trailing part flatter than the leading part.}
\label{fig_peaklrfs}
\end{figure}

\begin{figure*}
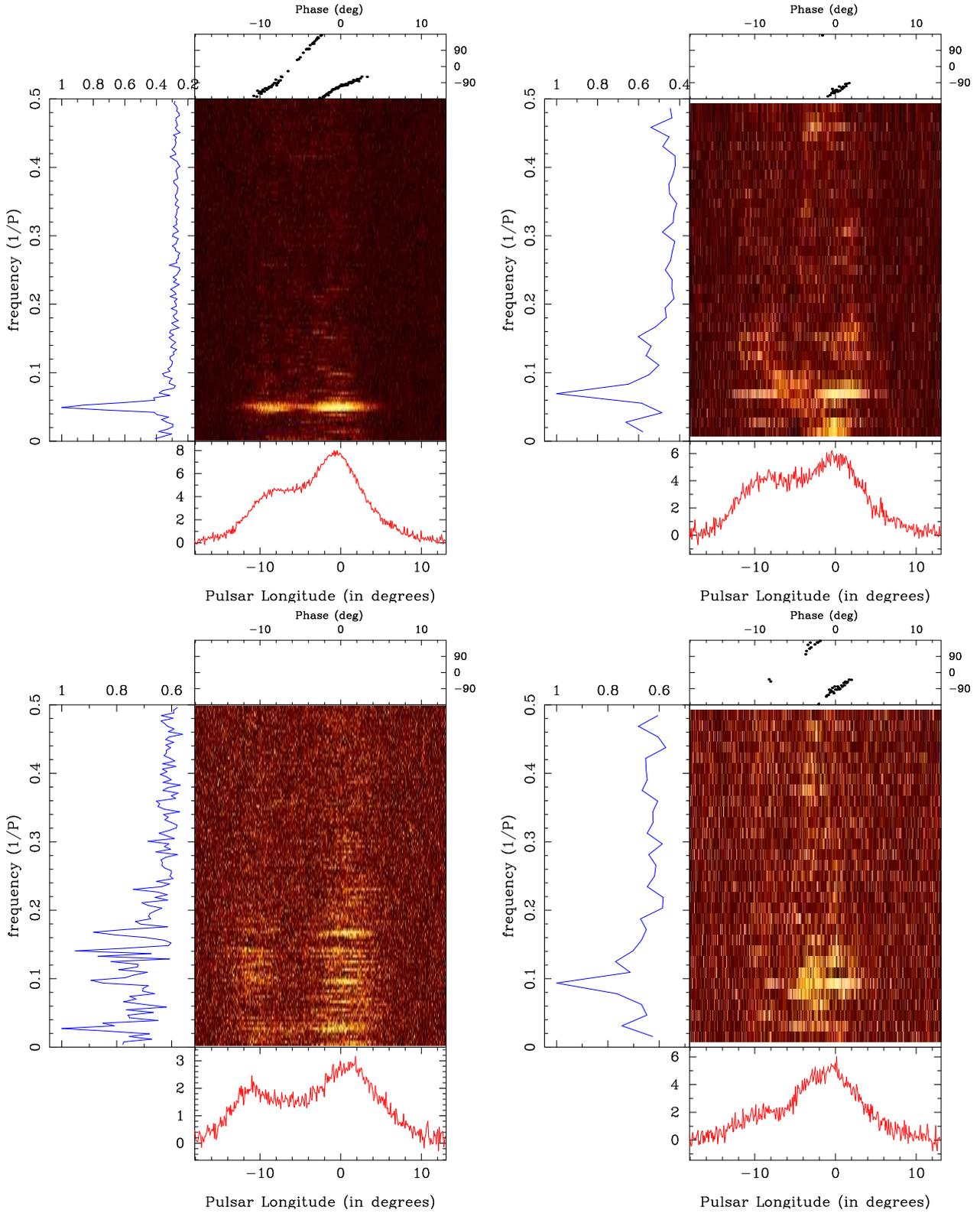

\begin{tabular}{@{}lr@{}}
{\mbox{\includegraphics[scale=0.42,angle=0.]{modeAlrfs.ps}}} &
{\mbox{\includegraphics[scale=0.42,angle=0.]{modeTlrfs.ps}}} \\
{\mbox{\includegraphics[scale=0.42,angle=0.]{modeBlrfs.ps}}} &
{\mbox{\includegraphics[scale=0.42,angle=0.]{modeClrfs.ps}}} \\
\end{tabular}
\caption{The figure shows the Longitude Resolved Fluctuation Spectra (LRFS) for
four different single pulse sequences corresponding to mode A (top left), the 
transitional drifting state before the onset of nulling in mode A (top right), 
the disorderly mode B (bottom left) and the drifting mode C (bottom right). 
The LRFSs show the drifting to be present throughout the pulse window 
in the first two cases but only near the trailing edge in mode C. In contrast 
no clear drifting peak is seen in mode B. However, a widish structure can be 
discerned below 0.2 cycles/$P$ indicating disordered subpulse motion during 
this mode. The drifting periodicities are 19.6$\pm$1.6 $P$ in mode A, 
14.3$\pm$1.8 $P$ during the transition to nulls in mode A and 10.7$\pm$1.1 $P$ 
in mode C which do not seem to be harmonically related to each other.}
\label{fig_lrfsmode}
\end{figure*}

We have carried out detailed measurements of the average drifting properties as 
well as in individual emission modes using the fluctuation spectral analysis. 
In figure \ref{fig_lrfsfull} we show the time evolution of the LRFS as detailed 
in \citet{bas18}. The typical FFT length for each time realisation of the LRFS
in these plots was 256 periods. The starting point was shifted by fifty periods
and the process was continued till the end of the observing duration. Each such 
realisation of the LRFS was averaged across the longitudes and plotted as a 
function of the starting period. The time averaged LRFS showed the presence of 
two distinct peaks, one around zero frequency which is associated with nulling,
and the second corresponding to the subpulse drifting in the most dominant mode
A. It should be noted that during one 256 period cycle, the pulsar is expected 
to make multiple transitions from mode A, with prominent drifting, to mode B, 
without any distinct drifting peak. This contributed to the peak amplitude 
being weaker and the peak more wider than expected from just mode A. The 
periodicities corresponding to the other drifting states were not seen as 
separate peaks in the average LRFS due to their low prevalence in the pulse 
sequence. However, they contributed to the wide structure adjacent to the peak 
frequency. The variations of the drifting peak across the pulse window are also
shown in figure \ref{fig_peaklrfs}. The peak amplitude (top window) showed a 
double peaked structure with the trailing part twice as high as the leading 
one. Additionally, the phase variations across the window (middle window) were 
not linear but were steeper towards the leading part of the profile and became 
flatter towards the trailing side. The phase variations were large amounting to 
about 600\degr across roughly 20\degr~variation in longitude. 

In addition we have also isolated single pulse sequences corresponding to the
different emission states and determined the fluctuation spectra as shown in 
figure \ref{fig_lrfsmode}. The figure shows four separate LRFS, the top left 
plot corresponding to mode A; the top right plot for a sequence during the 
transitional drifting state in mode A before the onset of nulling; the bottom
left plot shows the LRFS in mode B; and finally the bottom right plot 
corresponds to the pulse sequence in mode C. We have measured the drifting 
features in the three sequences where a clear frequency peak could be 
identified as reported in Table \ref{tabdrift}. The periodicity in Mode A was 
19.6$\pm$1.6 $P$ which changed to 14.3$\pm$1.8 $P$ before the onset of nulls. 
Mode C showed a different periodicity of 10.7$\pm$1.1. We do not see any clear 
harmonical relation between the different periodicities in this pulsar. 
The LRFS also showed the drifting to be present across the entire pulse profile
in mode A but only near the trailing part of the profile in mode C. The 
strength of the drifting peak was estimated using $S$ = $V_p$/FWHM, where $V_p$
was the peak height and FWHM the full width at half maximum. The $S$ factor was
strongest for the dominant drift mode A (61.8) and considerably weaker in mode 
C (16.5). We have also measured the average separation between the subpulses, 
$P_2$, in each of the drifting states using auto-correlation across the pulse 
longitude and subsequently averaging them for all relevant single pulses. The 
$P_2$ also varied in the three drift states with minimum separation of 
7.6$\pm$0.1\degr~in longitude for mode C. The corresponding value in mode A was
8.3$\pm$0.1\degr~which changed to 9.1$\pm$0.1\degr~in the transition state 
before nulling. The fluctuation spectra in the disorderly mode B also shows 
some wide signal below frequencies of 0.2 cy/$P$. However, there was no clear 
frequency peak seen in this emission state. At certain short intervals the
single pulse sequences during this mode also showed subpulse variations with 
periodicities seemingly less than mode A. This signifies that the pulse 
sequence during mode B had short bursts of drifting but no sustained ordered 
pattern. The different drifting periodicities and $P_2$ values reported here
differ from the drift mode estimates reported in \citet{nai17} for this 
pulsar. The different drift periodicities in this earlier work were estimated 
from average fluctuation spectra involving 256 periods. Our analysis showed 
that all emission states have much shorter durations and their behaviour would 
be diminished in these longer integration studies which may lead to erroneous 
measurements.

\section{Nulling}
\begin{figure}
\begin{center}
\includegraphics[scale=0.32,angle=-90.]{energyhist.ps} \\
\includegraphics[scale=0.47,angle=-90.]{nullenhist.ps} \\
\includegraphics[scale=0.47,angle=-90.]{burstlenhist.ps} \\
\includegraphics[scale=0.50,angle=0.]{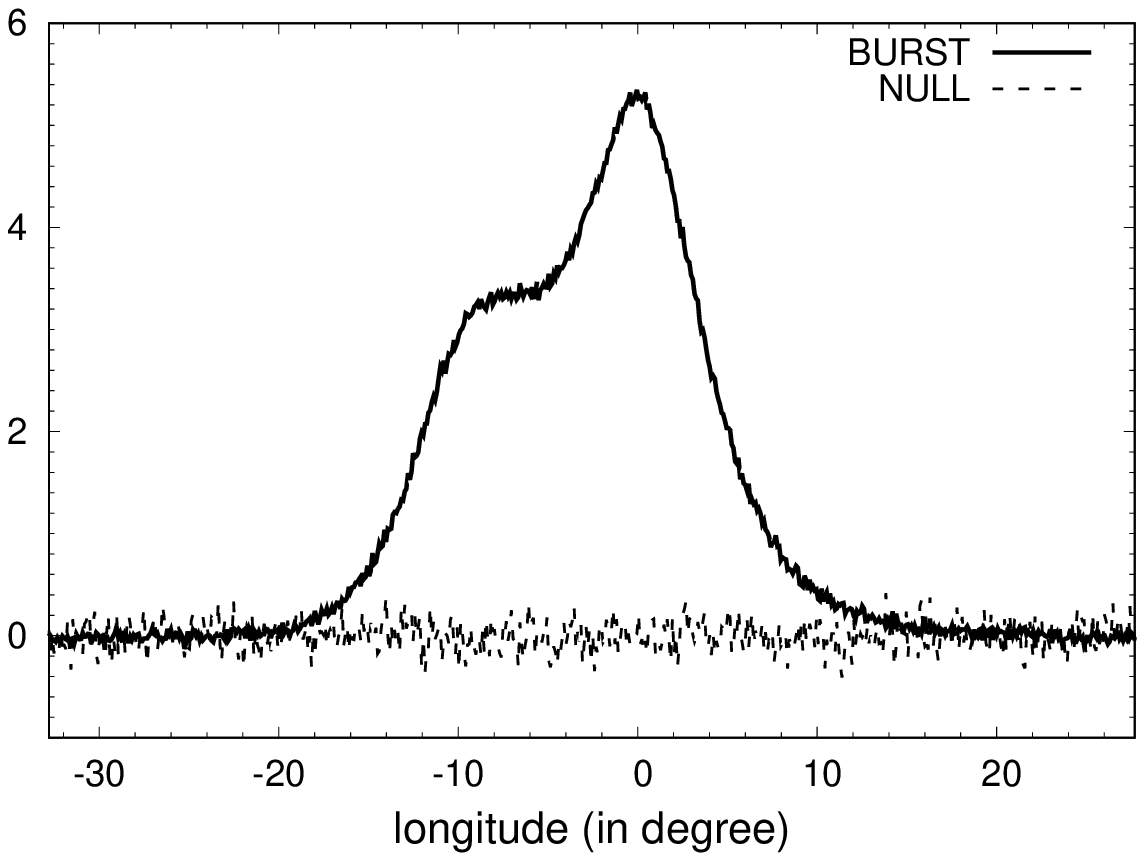} \\
\end{center}
\caption{The figure shows the different analyses to characterise the nulling in
the pulsar J1822$-$2256. The top panel shows the histograms corresponding to 
the average energy distribution, in arbitrary units, in the on-pulse window 
(blue) as well as the noise distribution from an off-pulse window (red), which 
resembles a Gaussian function centered around zero. The on-pulse energies show 
a peak around zero which correspond to the null pulses. The two central panels 
show the null length and burst length histograms for the pulse sequence. The 
nulls are of short duration and mostly last for one or two periods. Finally, 
the bottom plot shows the folded profiles for the null and burst pulses 
separately. The null profile is noise like and does not show any pulsed signal 
which validates our identification scheme for nulling.}
\label{fig_nullstat}
\end{figure}

\begin{figure}
\includegraphics[scale=0.43,angle=0.]{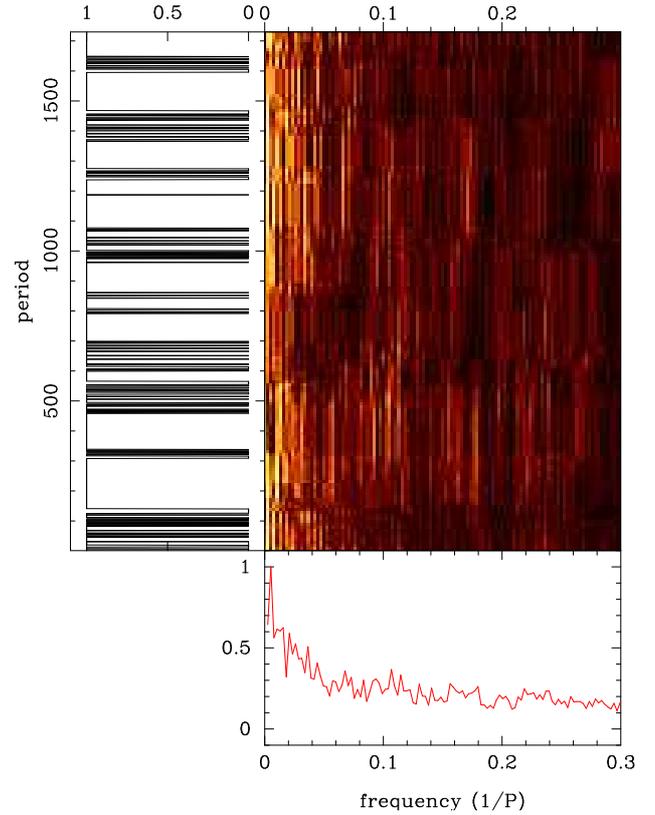} \\
\caption{The figure shows the zoomed in (along x-axis) FFT for the Null/Burst 
sequence where the nulls have been replaced with `0' and bursts with `1'. The 
nulling periodicity was estimated using an FFT length of 512 periods at a time.
The starting period was shifted by ten periods and the process was repeated to 
estimate the time variations in the nulling periodicity.}
\label{fig_nullfft}
\end{figure}

\begin{table}
\resizebox{\hsize}{!}{
\begin{minipage}{80mm}
\caption{Characterising the nulling in PSR J1822$-$2256.}
\centering
\begin{tabular}{cccccc}
\hline
 $N_P$ & $NF$ & $N_T$ & $\langle BL\rangle$ & $\langle NL\rangle$ & $P_N$ \\
   & (\%) &  &  &  & ($P$) \\
\hline
   &  &  &  &  &  \\
 5693  & 5.5$\pm$0.2 & 211 & 26.0 & 2.1 & 134$\pm$33 \\
   &  &  &  &  &  \\
\hline
\end{tabular}
\label{tabnull}
\end{minipage}
}
\end{table}

The nulling in the pulsar J1822$-$2256 was previously reported in \citet{bas17,
nai17}. \citet{bas17} suggested that the presence of the low frequency peak in 
the fluctuation spectra was a manifestation of periodic nulling which is only 
seen in around twenty pulsars. The presence of periodic nulling along with 
subpulse drifting in the same pulse sequence is even more rare and was 
previously reported in six other pulsars. This was a primary motivation for 
\citet{bas17} to suggest the two phenomena to have different physical origin. 
We followed the analysis schemes detailed in \citet{bas17} to estimate the 
nulling properties. The primary analyses to characterize nulling are shown in 
figure~\ref{fig_nullstat} which include the energy distributions in the 
on-pulse window as well as the off-pulse region (top panel), the null and burst
length histograms (middle panels) and the separate folded profiles of null and 
burst pulses (bottom plot). The details of nulling are also summarized in 
table~\ref{tabnull}. The nulling fraction ($NF$) was 5.5$\pm$0.2 percent which 
is consistent with the measurements of \citet{bas17}, which had shorter 
observing durations, but is different from the 10$\pm$2 percent reported in 
\citet{nai17}. We have measured 211 transitions ($N_T$) from the burst state to
the null state and vice versa. The null states were dominated by shorter 
duration nulls as seen in the null length histogram. The average null length 
($\langle NL\rangle$) was around two periods. The burst lengths on the other 
hand were of much longer durations reaching a maximum of around 300 consecutive
periods and an average length ($\langle BL\rangle$) of around 26 periods. 

The zero frequency structure in the average LRFS (Fig. \ref{fig_nullfft}) was 
likely associated with nulling. Given the long duration of observations we 
explored the possibility of resolving this periodicity by using longer duration
FFT studies and firmly associating the periodicity with the nulling process. We
used the null/burst sequence FFT as described in \citet{bas17}, where the nulls
were replaced with `0' and bursts with `1' to produce a time series of binary 
numbers. An FFT of this sequence was carried out to estimate the periodicity of
nulling. We have experimented with varying FFT lengths and finally used 512 
periods which enabled us to clearly separate the peak frequency from the zero 
boundary. The starting position was shifted by ten periods and the FFT was 
repeated till the end of the observing duration to determine the time 
variations as shown in figure~\ref{fig_nullfft}. A time averaged FFT was 
estimated and used to calculate the nulling periodicity. Our analysis clearly 
demonstrates that the low frequency structure seen in the LRFS corresponds to 
nulling. The corresponding periodicity ($P_N$) was calculated to be 
134$\pm$33$P$.

\section{Discussion}
\begin{figure*}
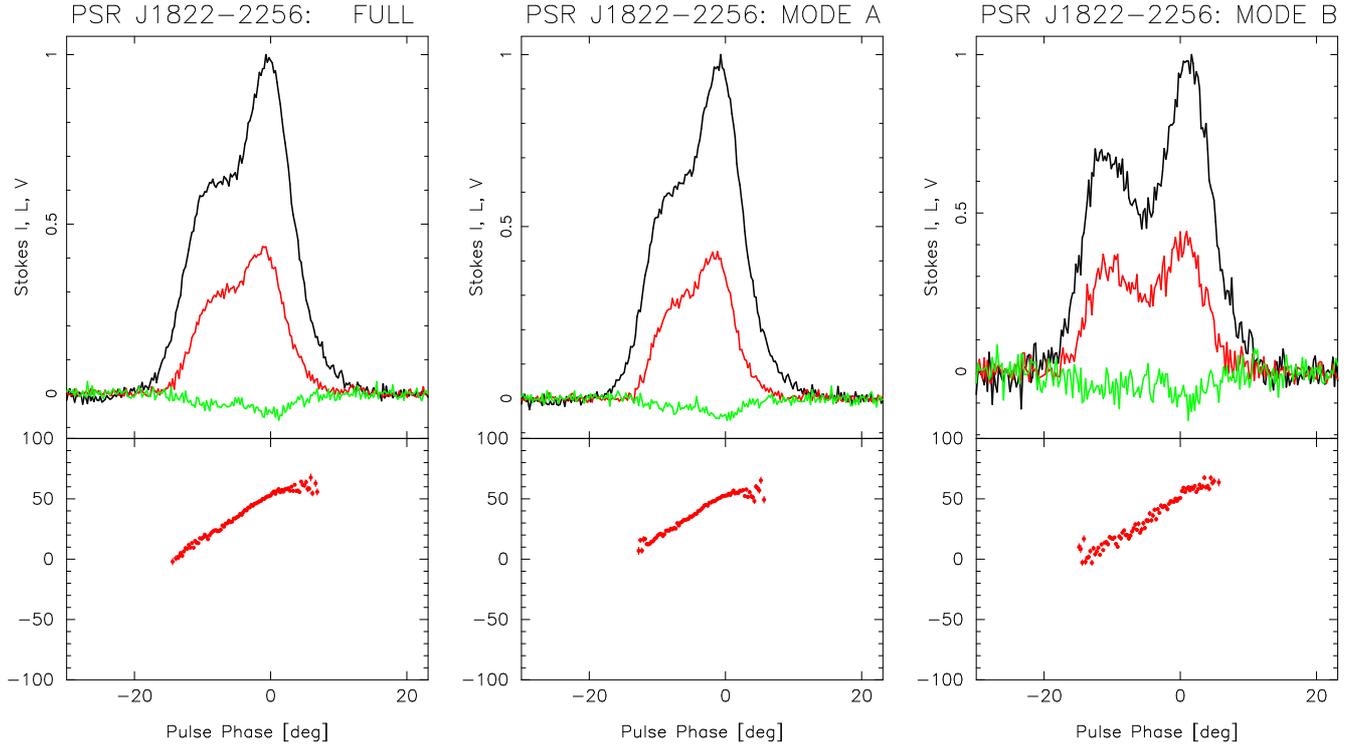

\begin{tabular}{@{}cr@{}cr@{}}
{\mbox{\includegraphics[scale=0.4,angle=0.]{J1822-2256_fullpol.ps}}} &
{\mbox{\includegraphics[scale=0.4,angle=0.]{J1822-2256_modeApol.ps}}} &
\hspace{10px}
{\mbox{\includegraphics[scale=0.4,angle=0.]{J1822-2256_modeBpol.ps}}} \\
\end{tabular}
\caption{The figure shows the average polarization properties of the pulsar 
J1822$-$2256 corresponding to the full observing session (left panel), Mode A 
(middle panel) and Mode B (right panel). The polarization information was
recorded in the shorter duration MSPES studies \citep{mit16}. We could only 
distinguish the two prominent modes A and B from these observations. The top 
window in each plot shows the average profile (black line) along with the 
linear polarization (L, red line) and the circular polarization (V, green 
line). The fraction of linear polarization varies slightly in the two modes. 
The bottom window in each figure shows the polarization position angle (PPA). 
The PPA is identical within measurement errors for all three cases.}
\label{fig_modepol}
\end{figure*}

\begin{table}
\resizebox{\hsize}{!}{
\begin{minipage}{80mm}
\caption{Emission Properties at different Modes.}
\centering
\begin{tabular}{ccccc}
\hline
 Mode & $W_{3\sigma}$ & $W_{10}$ & $\%L$ & $\%V$ \\
   & (\degr) & (\degr) &   &  \\
\hline
   &  &  &  &  \\
Full.  & 35.7$\pm$0.5 & 25.0$\pm$0.5 & 38.1$\pm$0.4 & -5.3$\pm$0.4 \\
   &  &  &  &  \\
Mode A & 35.4$\pm$0.5 & 22.7$\pm$0.5 & 35.6$\pm$0.4 & -4.2$\pm$0.4 \\
   &  &  &  &  \\
Mode B & 35.6$\pm$0.5 & 28.6$\pm$0.5 & 39.1$\pm$1.0 & -8.0$\pm$2.0 \\
   &  &  &  &  \\
Mode C & 25.3$\pm$0.5 & 22.3$\pm$0.5 & --- & --- \\
   &  &  &  &  \\
\hline
\end{tabular}
\label{tabavg}
\end{minipage}
}
\end{table}

\begin{figure}
\begin{center}
\includegraphics[scale=0.6,angle=0.]{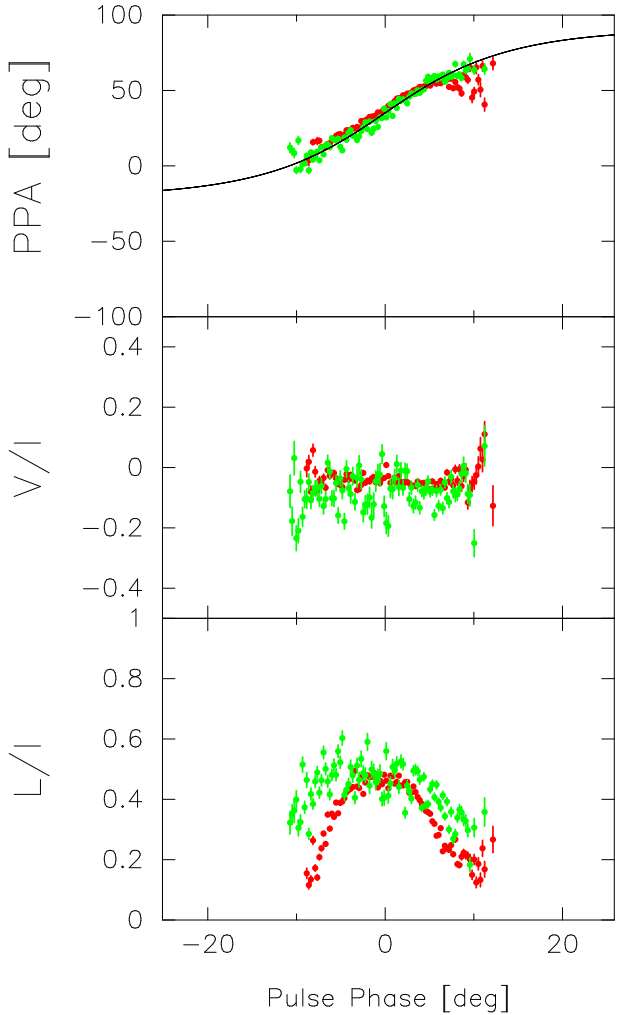}
\end{center}
\caption{The figure shows the comparison of the polarization properties across
the pulse window for the two emission modes A (in red) and B (in green). The 
top window shows the polarization position angle (PPA) in the two modes. A 
realization of the rotating vector model (RVM, black line) is also shown which
accurately fits the PPA in the two modes. The phase center of the RVM is 
located at -4.37\degr~longitude compared to the profile peak and we have 
shifted the plots to coincide with the phase center. The middle window shows 
the fractional circular polarization ($V/I$) at each pulse longitude. The 
bottom window shows the fractional linear polarization ($L/I$) at each pulse 
longitude.}
\label{fig_polcomp}
\end{figure}

\subsection{The Emission Region}
The subpulse drifting is a result of the dynamics of the sparking discharges in
the IAR characterised by large non-dipolar magnetic fields. The radio emission 
on the other hand originates at heights of few hundred kilometers above the 
stellar surface which are dominated by dipolar magnetic field lines. The 
presence of multiple drifting states and mode changing gives an unique 
opportunity to investigate if these changes have a corresponding effect on the 
emission process as well, primarily the emission altitudes. 

The characterisation of the pulsar geometry as well as the location of the 
radio emission require polarimetric observations \citep[see][for a review]{
mit17b}. The polarization position angle (PPA) across the pulse window shows a 
characteristic S-shape curve. The PPA is interpreted using the Rotating Vector 
Model \citep[RVM,][]{rad69} where the radiation is highly beamed and arises 
from regions of dipolar diverging field lines. The S-shaped curve is formed as 
the line of sight traverses across the diverging field lines. The pulsar 
geometry is characterised by $\alpha$, the angle between the rotation axis and 
the dipolar magnetic axis, and $\beta$, the angle between the rotation axis and 
the observers' line of sight. A relation between the steepest gradient (SG) 
point of the PPA and the pulsar geometry can be estimated using the RVM as SG = 
$\sin(\alpha))/\sin(\beta)$. It has been observed that correlations exist 
between the profile type and the shape of the PPA \citep{ran93}. For a highly 
central cut of the emission beam the SG points of the PPA traverse is large and
the profile has multiple components with core and conal emission. As the 
observer cuts the emission beam more tangentially the SG is less steep and the 
profile shape varies from a double to a single component. Detailed beam shape 
studies have revealed that the average emission beam comprises of nested cones 
around a central core emission region \citep{mit99}. 

The phase-modulated subpulse drifting, as seen in PSR J1822$-$2256, is usually
associated with conal profiles with shallow PPA traverses. The profile shapes
of these pulsars also show an evolution with frequency where low frequency 
double profiles usually become single component profiles at very high 
frequencies \citep{ran93}. Hence, average profiles at multiple frequencies are 
required to classify the profile type. We did not find any previous 
classification for the pulsar J1822$-$2256~in the literature. However, average 
profiles at multiple frequencies were available from an archival 
database\footnote{EPN pulsar Database}. Additionally, we also used the 
polarization observations from MSPES to estimate its emission properties. The 
left panel of figure \ref{fig_modepol} shows the polarization behaviour for the
average profile at 333 MHz. The PPA traverse is relatively shallow with 
estimated SG = 4.0 \degr/\degr. This is indicative of a tangential line of 
sight cut of the emission beam. The average profiles of this pulsar also show a
single component at higher frequencies (for example at 4.85 GHz). We conclude 
that the profile classification for this pulsar is consistent with conal single
type.

Next we proceed to compare the emission region in the different modes. The 
average profiles in the three primary modes (see figure \ref{fig_modeprof}) 
show very different shapes. However, comparison with the complete pulsar 
profile in each case (shown as dot dashed line in figure \ref{fig_modeprof}) 
suggests that the widths remain largely unchanged for the three modes. We have 
estimated the widths at three times the rms level of the baseline 
($W_{3\sigma}$) as well as the ten percent of the peak height ($W_{10}$) as 
shown in table \ref{tabavg}. The $W_{3\sigma}$ were identical for modes A and 
B. Due to higher baseline noise levels the estimated $W_{3\sigma}$ was lower 
for the profile in mode C. However, the $W_{10}$ in this case was once again 
identical to mode A. This signifies that in the different modes the emission 
regions were bounded by similar points along the open field lines. We could 
only identify the modes A and B in the MSPES observations and determined their 
average polarization behaviour as shown in figure \ref{fig_modepol} (middle and
left panel, respectively). Additionally, we have also carried out comparisons 
of the PPA traverse, the linear and circular polarization across the pulsar
profile for the two modes as shown in figure \ref{fig_polcomp} (mode A in red 
and mode B in green). The PPA traverses (top panel) were identical for the two 
modes with the same RVM (black line) fitting both of them. The RVM fit shown in
the figure corresponds to $\alpha$ = 16.2\degr~and $\beta$ = 4.0\degr\footnote{Fitting the RVM to the PPA gives estimates of the angles $\alpha$ and $\beta$.
However, these estimates are highly correlated and do not give good constrains
on the geometry. Rather the SG point is significantly better determined in the 
PPA traverse.}. Mode B seemed to have slightly higher linear polarization 
towards the leading and trailing part of the profile but no discernible 
difference in circular polarization could be seen in the two modes. Such slight
variations in fractional polarization can be associated with changes in plasma 
inhomogeneity \citep{mel14}. These analyses show that during the mode changes 
the emission region remains largely unaffected and the emission continues to 
arise from the same heights.

The radio emission arises due to non linear plasma processes where charged 
bunches (relativistic solitons) excite coherent curvature radiation in curved 
magnetic fields. The emission from a large number of such charged bunches adds 
up incoherently to give the observed radio intensity \citep{ass98,mel00,gil04}.
In such a model the characteristic frequency of emission ($\nu_c$) is given as 
$\nu_c \sim \gamma^3 c/ \rho_c$, where $\gamma$ is the Lorentz factor of the 
radiating plasma and $\rho_c$ is the underlying radius of curvature of the 
magnetic field. As argued above that during mode transitions the geometry 
remains unchanged which implies that $\rho_c$ across the emission region is 
also unchanged for the three modes. This further suggests that in order to get
the same $\nu_c$, the $\gamma$ of the radiating plasma also needs to be similar
in the different modes. The power of the radio emission in this model is $P 
\propto F(Q) \gamma^4/\rho_c^2$, where $F(Q)$ has the dimension of charge 
squared and is a complex function of the plasma parameters. The observations 
show that despite the emission geometry remaining same across the profile, the 
emitted power at different longitudes varies in the three modes. This can only 
arise due to variation in the $F(Q)$ which depends on the changes in the plasma
parameter in a complicated manner. Similar conclusions have also been drawn for
the mode changing pulsar PSR B0329+54 where observations revealed that the 
locations of the radio emission were similar for the different modes 
\citep[][Brinkman, Mitra \& Rankin, 2017, private communication]{bar82}. 
The plasma changes are likely driven by variations in the IAR where they are 
generated. In summary, our analyses reveal that the variations seen during mode
changing are unlikely to be affected by changes in the emission region and are 
possibly driven by the complex plasma processes which take place during their
generation in the IAR.

\subsection{Variations in Subpulse Drifting during Mode Changing}
The presence of multiple drifting states in addition to a disorderly mode and 
periodic nulling in the same pulsar gives an unique opportunity to further 
investigate the physical conditions in the magnetosphere. As argued in the 
previous section the emission regions for different modes are similar and the 
mode changing is driven by variations in the IAR. In this regard the presence 
of different drifting properties is useful to better understand the conditions
in the IAR. The sparks in the IAR responsible for the generation of the 
outflowing plasma have typical timescales ranging from hundreds of nanoseconds 
to microseconds \citep[][hereafter RS75]{rud75}. These are much shorter than 
the drifting periodicities which represent average behaviour of the conditions 
in the IAR. The drifting periodicity can be can be estimated as $P_3 = d/v_d$, 
where $d$ is the average separation between two consecutive sparks and $v_d$ 
the drift velocity of the sparks. In the IAR $v_d$ can be further expressed as 
$v_d = (\Delta E/B) c$, where $\Delta E$ is the change in the electric field in
the gap during the sparking process, $B$ the magnetic field in the IAR and $c$ 
the speed of light. It is difficult to see how the magnetic field associated 
with the star can change at these timescales since this will result in large 
scale reorientation of the current flow in the pulsar circuit \citep{spi11}. 
Hence, the different $P_3$ values can be attributed to either changes in $d$ or
$\Delta E$ or combinations of both, i.e. on the term $d/\Delta E$. The 
separation between the sparks can be approximately related to $P_2$ and the 
pulsar geometry as 
$d \propto 2\pi\left(\frac{P_2}{360\degr}\right)\sqrt{\frac{R_S^3}{R_E}}\sin(\alpha+\beta)~b^{0.5}$. 
Here, $R_S$ is the radius of the neutron star, $R_E$ the height of the emission
region and $b$ the scaling factor between dipolar component and the non-dipolar
field in the IAR. As reported earlier the three drifting states have different 
$P_2$ values. We have argued above that the emission geometry and $R_E$ in the 
three modes are similar. This implies that the $P_2$ variations in the emission 
region are indicative of changes in the spacing between sparks. Hence, the 
number of sparks in the IAR are different in each emission mode. According to 
RS75 the sparking process is governed by the energy of background $\gamma$-ray 
photons and the IAR electric field $\Delta E$. A change in $P_2$ is a direct 
indication that the $\Delta E$ in the gap is also changing. There are however 
no provisions for such changes in the steady state models like RS75.

As noted earlier mode changing, nulling and drifting phenomena occur on 
timescales which are significantly larger than the dynamics of the plasma 
formation. This has also been recognized in earlier observations of mode 
changing where external mechanisms were expected to change the plasma flow 
during the changes \citep{bar82}. In several recent studies \citep{bas16,bas17,
mit17,raj17} the mode changing and nulling phenomena have been reported to have
a periodic/quasi-periodic nature. It was suggested by \citet{bas17} that an 
external triggering mechanism in the larger magnetosphere is required to drive 
the periodic changes in the plasma generation process of the IAR. In this work 
we show the presence of periodic nulling as well as regular mode changing in 
the same system. The nulling periodicity is 134$\pm$33 $P$. The pulsar 
primarily exists in the two modes A and B. We found the average duration of 
mode A to be 82 periods and that of mode B to be 68 periods. This means that on
average the pulsar comes back to its initial mode at a timescale of 150 (82+68)
$P$, which is comparable to the nulling periodicity. This prompts the 
interesting possibility that the triggering mechanism proposed earlier for the 
periodic nulling phenomenon also induces the mode changing in this pulsar. 
However, the physical origin of the triggering mechanism is unknown. 

The different drifting states associated with the emission modes have been 
investigated in the past using the carousel model. The significant frequency 
evolution of profile widths in J1822$-$2956 suggests the presence of outer 
cone which has been argued to have a beam radius, $\rho$ = 5.75\degr/$P^{0.5} 
\approx$ 4.2\degr \citep{ran93}. The number of circulating beamlets in the 
different drifting states can be estimated using $\rho$ and the corresponding 
$P_2$ values \citep{des01}. It was argued by \citet{ran13} that in the pulsar 
B1918+19 the carousel circulation time was constant across the different 
drifting modes. The number of subbeams making up the circulation pattern 
changed in the different drifting states leading to the difference in the 
measured $P_3$. It is possible to extend these analysis schemes for the 
drifting modes of J1822$-$2956 as well, but recent observations have raised
questions about the applicability of carousel model in pulsars \citep{bas16,
mit17}. The preponderance of short duration nulls known as `pseudo nulls', seen
periodically, as reported here for the pulsar J1822$-$2956 has been used as 
additional justification for the carousel model \citep{her07,her09}. The pseudo
nulls are associated with the line of sight passes across the empty regions 
within the rotating subbeam system. However, the presence of pseudo nulls 
reported in core dominated pulsars challenges this interpretation \citep{bas17}.

Another possibility for the mode changing and nulling has been suggested by 
\citet{tim10}. In this model the transitions to the emission states are 
governed by corresponding variations in the global magnetospheric current flow 
which can change the extent of the open field line regions. A possible 
observational indication is difference in the profile widths and geometries in 
the different modes. However, similar widths of the profiles as well as the 
geometries for the three modes argue against the model to be applicable in this
pulsar.

\section{Summary}
We have carried out detailed analyses of the single pulse dynamics in the 
pulsar J1822$-$2256. We have identified several distinct emission states and 
have categorized them as three different modes based on their relative 
abundances. The most dominant mode seen for around 45 percent times was 
identified as mode A. This mode consisted of prominent drift bands seen across 
the entire profile from the leading to the trailing edge. Mode B was seen for 
around 38 percent of the observing duration and was more disorderly without the
presence of any clear subpulse drifting. A third mode C is comprised of a 
second drift mode with periodicity lower than mode A. The emission state was 
seen for a single interval lasting for around 200 consecutive periods and was 
more bright towards the trailing part of the profile. In addition to the three 
distinct modes the pulsar also showed the presence of nulling throughout the 
observing durations. The nulling was present in all the states but was more 
frequent during the disorderly mode B. In some cases the drifting behaviour in 
mode A just before the onset of nulls changed, showing a periodicity different 
from both modes A and C. We have also shown the nulling to exhibit periodicity 
which is much larger than the drifting periodicities. The radio emission in 
this pulsar is unique owing to the diversity of single pulse phenomena seen in 
the same system. We have shown that the emission region remains unchanged in 
the three modes and the mode changing is likely driven by variation in the 
plasma generation process. The regular change in the emission process cannot be
explained using a conventional steady state model of plasma generation. 
Additional triggering mechanism from the pulsar magnetosphere would be required
to change the plasma processes periodically/quasi periodically. This requires 
more detailed modelling which can then be constrained using the detailed 
observational inputs which have been gathered in recent works. Recent studies 
also show the radio mode changes have corresponding changes in the X-ray flux 
\citep{her13}. This also indicates a change in the plasma generation process 
during mode changing.

\section*{Acknowledgments}
We thank the referee Joanna Rankin for her comments which helped to improve the
paper. We thank George I. Melikidze for discussions about subpulse drifting. We
thank the staff of the GMRT who have made these observations possible. The GMRT
is run by the National Centre for Radio Astrophysics of the Tata Institute of 
Fundamental Research.

\end{document}